# Building a Reusable and Extensible Automatic Compiler Infrastructure for Reconfigurable Devices

Zhenya Zang[†, *]
[†]Codeplay Software Ltd.
[*]School of Computing, Engineering and Build Environment,
Edinburgh Napier University
Edinburgh, United Kindom
zhenya.zang@codeplay.com

Uwe Dolinsky[†]
[†]Codeplay Software Ltd.
Edinburgh, United Kindom
uwe@codeplay.com

Pietro Ghiglio[†]
[†]Codeplay Software Ltd.
Edinburgh, United Kindom
pietro.ghiglio@codeplay.com

Stefano Cherubin[*]
[*]School of Computing, Engineering and Build Environment
Edinburgh Napier University
Edinburgh, United Kindom
s.cherubin@napier.ac.uk

Mehdi Goli[†]
[†]Codeplay Software Ltd.
Edinburgh, United Kindom
mehdi.goli@codeplay.com

Shufan Yang[*]
[*]School of Computing, Engineering and Build Environment
Edinburgh Napier University
Edinburgh, United Kindom
s.yang@napier.ac.uk

*Abstract*— **Multi-Level Intermediate Representation (MLIR) is gaining increasing attention in reconfigurable hardware communities due to its capability to represent various abstract levels for software compilers. This project aims to be the first to provide an end-to-end framework that leverages open-source, cross-platform compilation technology to generate MLIR from SYCL. Additionally, it aims to explore a lowering pipeline that converts MLIR to RTL using open-source hardware intermediate representation (IR) and compilers. Furthermore, it aims to couple the generated hardware module with the host CPU using vendor-specific crossbars. Our preliminary results demonstrated the feasibility of lowering customized MLIR to RTL, thus paving the way for an end-to-end compilation.**

*Keywords—Open-source FPGA compiler infrastructure, Multi-Level Intermediate Representation (MLIR), Hardware description language (HDL)*

## I. Introduction

Intermediate representation (IR) and compilers of FPGA bridge the gap between software programming languages and the hardware domain. A novel HDL [1] was reported to simplify hardware design using its compiler and Medium-Level IR. Besides, open-source, software-defined HDL and compilers become emerging technologies for FPGAs and application-specific integrated circuits (ASIC). Chisel [2] is successful in modeling digital circuits using a high-level, objective-oriented language and has launched a high-performance, open-source CPU [3], and ASIC [4]. FIRRTL [5] is Chisel's IR and can be lowered to hardware primitives, i.e., combinational and sequential logic, and memories. Despite Chisel's agility for fast RTL prototyping, the program compilation for CPU and GPU has not been well-developed. An HDL that can target CPUs, GPUs, and FPGAs is demanding. Equipped with powerful compilers, SYCL has been adopted to accelerate computing for CPUs and GPUs. Herein, we adopt SYCL and DPC++ [6] as the front-end to generate Multi-Level Intermediate Representation (MLIR) [7]. In essence, MLIR aims to provide a common representation for different levels of abstraction, including high-level algorithmic descriptions. However, mapping these high-level abstractions to efficient reconfigurable hardware implementations is challenging. This involves translating the MLIR representation to RTL and fine-grained control over hardware primitives. We have explored a lowering pipeline for converting MLIR to RTL by using open-source circuit compilers CIRCT [8] and Calyx [9] to generate RTL. Our work has already led to several fixes and improvements[1] in the Calyx toolchain. The ultimate objective of the project is to encapsulate all the compilation and lowering processes to form an automated end-to-end framework for reconfigurable heterogenous platforms.

## II. On-going Work to Date

### A. Framework architecture

The sketch of our compilation framework is shown in Fig. 1. The framework begins with input SYCL code, which is processed by DPC++ to generate the corresponding MLIR compatible with CIRCT. The output Calyx dialects produced by CIRCT are then passed to Calyx to generate System Verilog. Using established EDA tools, we can acquire synthesized and simulation results. Additionally customized hardware drivers must be written to facilitate the booting of the heterogeneous platform on an FPGA.

---

[1] https://github.com/cucapra/calyx/issues/1470

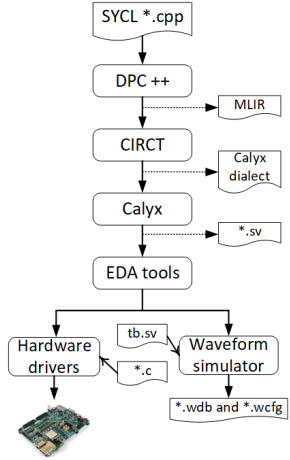

Fig. 1. End-to-end pipeline mapping SYCL code to System Verilog, and silumation path. Dashed lines denotes generate files.

## B. Preliminary Results

We selected general matrix multiplication (GEMM) as our case study to evaluate the performance in terms of hardware utilization and time consumption. The preliminary results demonstrated that our approach successfully generates accurate output matrices from MLIR and achieves improved speeds by applying for-loop unrolling to the inner for-loop. We have encapsulated the lowering pipelines using a script, enabling the generation of a hardware module wrapped with an AXI-full interface and a synthesizable intellectual property (IP) core. This encapsulation facilitates seamless integration of heterogeneous platforms with CPUs in the future.

The generated System Verilog was simulated and synthesized using Vivado 2020.2 with a ZCU106 evaluation board. Consumed clock cycles of the two versions of GEMM are presented in TABLE I, where GEMM with inner-unrolled for-loop yields fewer clock cycles. The corresponding hardware consumption is illustrated in Fig. 3 (a) and (b). GEMM with a nested for-loop utilized time division multiplexing, allowing the reuse of data paths and DSPs. However, in the flattened for-loop in Fig. 3 (b), the hardware consumption is directly proportional to the size of matrix.

## III. CONCLUSIONS AND FUTURE WORK

Our ongoing work has successfully incorporated the input MLIR into the open-source HDL compilers prompting various improvements, simulated the results, and evaluated the performance. The future work is threefold.

1) We will integrate DPC++ as the front-end of the framework, enabling the conversion of SYCL kernels into hardware via MLIR.

2) Since the Calyx tool is tailored for the Xilinx Runtime Library (XRT), which is dependent on Xilinx platforms, we will replace XRT with a standalone platform to extend the application on other lightweight, low-power hardware.

3) We plan to develop more complex application algorithms based on GEMM, such as tensor operations for machine learning algorithms.

The work is funded by a knowledge transfer partnership (reference 13191) from Innovate UK with the grant number 10027750. Shufan Yang is supported by the SHED project RAEng (IF2223-172).

```
1  func.func @mlir_funcSYCL_class_mxm_kernel(%arg0: memref<1024xi32, 1>,
2                                            %arg1: memref<1024xi32, 1>,
3                                            %arg2: memref<1024xi32, 1>) {
4    affine.for %arg3 = 0 to 32 {
5      affine.for %arg4 = 0 to 32 {
6        %c0_i32 = arith.constant 0 : i32
7        %1 = affine.for %arg5 = 0 to 32 iter_args(%arg6 = %c0_i32) -> (i32) {
8          %2 = affine.load %arg2[%arg5 * 32 + %arg4] : memref<1024xi32, 1>
9          %3 = affine.load %arg1[%arg5 + %arg3 * 32] : memref<1024xi32, 1>
10         %4 = arith.muli %2, %3 : i32
11         %5 = arith.addi %4, %arg6 : i32
12         affine.yield %5 : i32
13       }
14       affine.store %1, %arg0[%arg4 + %arg3 * 32] : memref<1024xi32, 1>
15     }
16   }
17   return
18 }
```

Fig. 2. GEMM MLIR with 32×32 input matrix size and nested for-loops

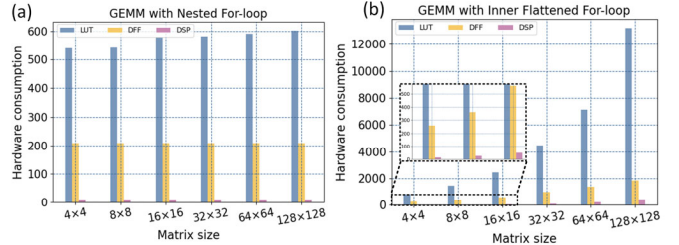

Fig. 3. (a). Hardware consumption of nested and (b) inner-flattened for-loops of GEMM MLIRs with different size.

TABLE I. CONSUMED CLOCK CYCLES (1NS PER CYCLE) OF COMPUTING GEMM WITH DIFFERENT SIZES

|  | 4×4 | 8×8 | 16×16 | 32×32 | 64×64 | 128×128 |
|---|---|---|---|---|---|---|
| Nested for-loop | 1,498 | 10,762 | 81,802 | 867,594 | 5,042,698 | 38,324,504 |
| Inner Flattened for-loop | 1,114 | 7,946 | 60,298 | 470,282 | 3,527,115 | 26,806,047 |

## ACKNOWLEDGEMENT

We would like to acknowledge Calyx's developer for answering technical questions on GitHub.